\documentclass{elsart5p}%

\usepackage{graphics}
\usepackage{graphicx}
\usepackage{amssymb}%
\usepackage{amsmath}%
\usepackage{amsfonts}
\begin{document}
\begin{frontmatter}
% Title, authors and addresses
% use the thanksref command within \title, \author or \address for footnotes;
% use the corauthref command within \author for corresponding author footnotes;
% use the ead command for the email address,
% and the form \ead[url] for the home page:
% \title{Title\thanksref{label1}}
% \thanks[label1]{}
% \author{Name\corauthref{cor1}\thanksref{label2}}
% \ead{email address}
% \ead[url]{home page}
% \thanks[label2]{}
% \corauth[cor1]{}
% \address{Address\thanksref{label3}}
% \thanks[label3]{}
\title{Disorder screening near the Mott-Anderson transition}
%
% use optional labels to link authors explicitly to addresses:
% \author[label1,label2]{}
% \address[label1]{}
% \address[label2]{}
\author[AA,BB]{M. C. O. Aguiar\corauthref{M. C. O. Aguiar}},
\ead{aguiar@fisica.ufmg.br}
\author[CC]{V. Dobrosavljevi\'{c}},
\author[BB]{E. Abrahams},
\author[BB]{G. Kotliar}
\address[AA]{Departamento de F\'{\i}sica, Universidade Federal de Minas Gerais,
Av. Ant\^onio Carlos, 6627, Belo Horizonte, MG, Brazil}
\address[BB]{Center for Materials Theory, Serin Physics Laboratory, Rutgers
University, 136 Frelinghuysen Road, Piscataway, New Jersey 08854, USA}
\address[CC]{Department of Physics and National High Magnetic Field
Laboratory, Florida State University, Tallahassee, FL 32306, USA}
\corauth[M. C. O. Aguiar]{Corresponding author. Tel: +55 31 3499-5671 fax:
+55 31 3499-5600}
\begin{abstract}
% Text of abstract
Correlation-driven screening of disorder is studied within the
typical-medium dynamical mean-field theory (TMT-DMFT) of the
Mott-Anderson transition. In the strongly correlated regime, the
site energies $\varepsilon_R^{i}$ characterizing the effective
disorder potential are strongly renormalized due to the phenomenon
of Kondo pinning. This effect produces very strong screening when
the interaction $U$ is stronger then disorder $W$, but applies
only to a fraction of the sites in the opposite limit ($U<W$).
\end{abstract}
\end{frontmatter}

%main text

\textit{Introduction} -
%\label{}
Theories that are able to capture both the Mott~\cite{mott74} and
the Anderson~\cite{anderson58} mechanisms for electron
localization have remained elusive despite many years of effort.
An attractive approach to this difficult problem has recently been
proposed by combining the dynamical mean-field theory
(DMFT)~\cite{dmftrev} of the Mott transition, and the typical
medium theory (TMT)~\cite{tmt} of Anderson localization. This new
formulation of the Mott-Anderson problem has been explored in
recent work by Vollhardt and collaborators \cite{vollhardt} using
numerical renormalization-group methods, but the precise mechanism
for the critical behavior of this model remains to be elucidated.
Here we examine the mechanism for disorder screening within this
theory, which explains several aspects of the results found in
Ref. \cite{vollhardt}

Within TMT-DMFT, a lattice problem is mapped onto an ensemble of
single-impurities problems, which are embedded in a
self-consistently determined bath. Recent work of
Ref.~\cite{impurity} examined the behavior of a collection of
single-impurity models in the situation where the bath seen by the
impurities was chosen to mimic the approach to the Mott-Anderson
transition. In this work, the impurity quasiparticle weight $Z_i$
was shown to present a scaling behavior as a function of the
on-site energy $\varepsilon_{i}$ and the distance $t$ to the
transition. These findings, however, are not sufficient to address
the disorder screening behavior of the model, which requires the
description of the renormalized disorder potential. In this paper,
we demonstrate that a scaling procedure similar to that presented
in Ref.~\cite{impurity} can also be carried on for the
renormalized energy $\varepsilon_{R}^{i}$.

%continue your text, include further sections or subsections if needed

\textit{Renormalization of the disorder potential} - We consider a
collection of Anderson impurity models~\cite{impurity} with
on-site repulsion $U$, on-site energies $\varepsilon_{i}$, and the
total spectral weight $t$ of the cavity field. Without loss of
generality~\cite{impurity}, we consider a featureless model bath
with non-vanishing spectral weight for $-t/2<\omega<t/2$ and zero
otherwise. Our goal is to describe the statistics of the
renormalized site energies as the metal-insulator transition is
approached, corresponding to  $t\rightarrow0$ within the TMT-DMFT
scheme.

\begin{figure}[ptb]
\begin{center}
\includegraphics[width=0.4\textwidth]{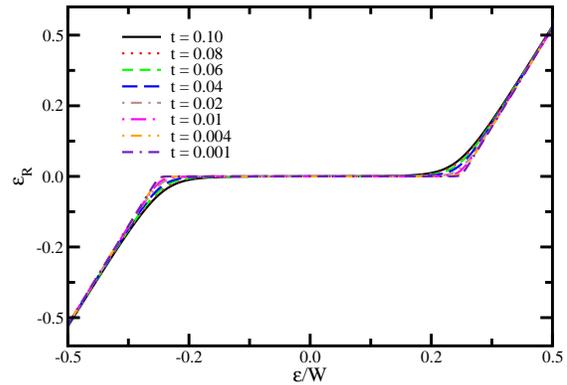}
\end{center}
\caption{Renormalized energy $\varepsilon_{R}$ as a function of
the on-site energy~$\varepsilon$ for a collection of
single-impurity problems close to the Mott-Anderson transition ($t
\rightarrow0$). The parameters used were $U=1.75$
and $W=2.8$.}%
\label{fig1}%
\end{figure}

The impurity models were solved at zero temperature using the SB4
method~\cite{sb4, impurity}, which provides a parametrization of
the low-energy (quasiparticle) part of the local Green's function,
given by
\begin{equation}
G_{i}(\omega_{n})=\frac{Z_{i}}{i\omega_{n}-\varepsilon_{R}^{i}-Z_{i}%
\Delta(\omega_{n})}.
\end{equation}
Here $Z_{i}$ is the local quasiparticle weight and
$\varepsilon_{R}^{i}$ is the renormalized site energy. The details
of the calculations mirror those of Ref.~\cite{impurity}.

The results for the renormalized energy $\varepsilon_{R}^{i}$ as a
function of $-W/2 < \varepsilon_{i}< W/2$, in the vicinity of the
Mott-Anderson transition ($t \rightarrow0$), are shown in
Fig.~\ref{fig1}. As in Ref.~\cite{impurity}, we find two-fluid
behavior, where sites with $|\varepsilon_{i}|<U/2$ turn into local
magnetic moments, corresponding to  ``Kondo
pinning''~\cite{screening} $\varepsilon_{R}^{i} \rightarrow0$.
For the remaining sites, $\varepsilon_{R}^{i} \rightarrow
\varepsilon_{i} + U/2$ or $\varepsilon_{R}^{i}
\rightarrow\varepsilon_{i} - U/2$, as they become, respectively,
doubly occupied (those with $\varepsilon _{i}<-U/2$) or empty
(those with $\varepsilon_{i}>U/2$). We should emphasize that such
two-fluid behavior thus emerges only for sufficiently strong
disorder, such that $U<W$. Otherwise all sites turn into local
magnetic moments, and the Mott transition for moderate disorder
retains a character similar to that found within the standard DMFT
approach~\cite{screening}.

\begin{figure}[ptb]
\begin{center}
\includegraphics[width=0.4\textwidth]{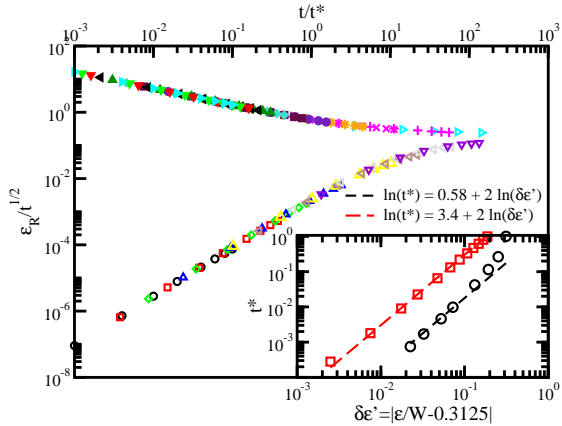}
\end{center}
\caption{Scaled renormalized energy $\varepsilon_{R}/t^{1/2}$ as a
function of $t/t^{\ast}(\delta\varepsilon)$ showing that the
results for different (and positive) $\varepsilon$ can be
collapsed onto a single scaling function with two branches.
Different symbols correspond to different $\varepsilon$; the upper
(bottom) branch contains results for $\varepsilon>U/2$
($\varepsilon <U/2$). The inset shows the scaling parameter
$t^{\ast}$ as a function of $|\varepsilon/W-0.3125|$ for the upper
(squares) and bottom (circles)
branches. The parameters used were $U=1.75$ and $W=2.8$.}%
\label{fig2}%
\end{figure}

\textit{Scaling analysis} - These results can alternatively be
presented in a scaling form, as shown in Fig.~\ref{fig2}. Here, we
show that it is possible to collapse the family of curves
$\varepsilon _{R}(t,\delta\varepsilon) /t^{0.5}$, where
$\delta\varepsilon=\left( \varepsilon_{i}-\varepsilon^{\ast}
\right)  /\varepsilon^{\ast}$ and $\varepsilon^{\ast} = U/2$, onto
a single universal scaling function
$\varepsilon_{R}(t,\delta\varepsilon)/t^{0.5}=f[t/t^{\ast}(\delta
\varepsilon)]$ with two branches, one for $\varepsilon_{i} <
\varepsilon ^{\ast}$ and other for $\varepsilon_{i} >
\varepsilon^{\ast}$. In agreement with Ref.~\cite{impurity} (inset
of Fig.~\ref{fig2}), the crossover scale
$t^{\ast}(\delta\varepsilon) \sim|\delta\varepsilon|^{\phi}$, with
exponent $\phi=2$. In the limit $t \rightarrow0$, we find that the
branch corresponding to $\varepsilon_{i} < \varepsilon ^{\ast}$
has the asymptotic form  $f(x) \sim x^{3/2}$ (here
$x=t/t^{\ast}(\delta\varepsilon)$), corresponding to
$\varepsilon_{R}(t) \sim t^{2}$. Similarly,  for $\varepsilon_{i}
> \varepsilon ^{\ast}$, $f(x) \sim x^{-1/2}$ corresponding to
$\varepsilon_{R}(t) \sim$ constant.
For $x\gg1$ the two branches merge, viz. $f(x)\sim A \pm B^{\pm}
x^{-0.5}$.

\begin{figure}[ptb]
\begin{center}
\includegraphics[width=0.4\textwidth]{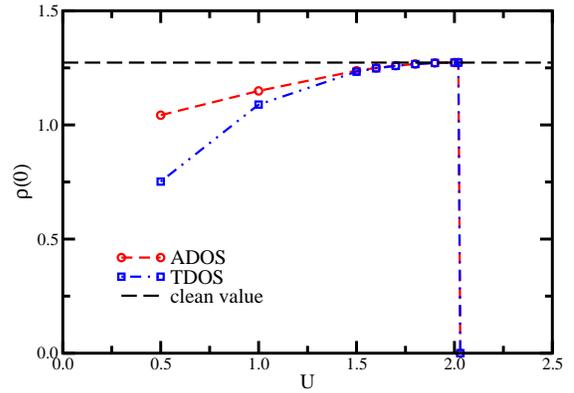}
\end{center}
\caption{Arithmetic and geometric density-of-states (ADOS and TDOS,
respectively) at the Fermi level as a function of $U$, for $W=1.5$, when the
TMT-DMFT self-consistent loop is performed.}%
\label{fig3}%
\end{figure}

\textit{Disorder screening} - Within TMT-DMFT, the Anderson
localization effects are manifested by the reduction of the {\em
typical} density of states (TDOS), since the (algebraic) average
(ADOS) remains finite even in an Anderson insulator.  When
disorder is strongly screened due to the correlation effects, the
two quantities should not differ much, as illustrated by the
results of Fig.~\ref{fig3}. Here we present the results of the
fully self-consistent solution, as the Mott-like transition is
approached by increasing $U$ for $W=1.5$. Close to the transition,
both averages approach the clean limit (dashed line), indicating a
strong screening effect. These results are consistent with those
found in the numerical renormalization group solution of the
TMT-DMFT equation of Ref.~\cite{vollhardt}.

As discussed above, strong disorder screening is expected near the
Mott-like transition ($U>W$), which indeed corresponds to the
mechanism responsible for the results in Fig.~\ref{fig3}.
When the transition is approached at strong disorder ($U<W$) (not
shown), strong screening effects are found only for a fractions of
the sites (i.e of the volume of the sample), indicating different
critical behavior at the Mott-Anderson transition. The details of the
critical behavior in this case will be discussed elsewhere.

\textit{Acknowledgements} - This work was partially supported by
%FUNDEP/UFMG (M.C.O.A.) and 
NSF grants DMR-0312495 (M.C.O.A.),
DMR-0234215 and DMR-0542026 (V.D.) and DMR-0096462 (G.K.).


\begin{thebibliography}{9}                                                                                                %
%\bibitem{Paper1} A. Author1, Adv. Phys. {\bf 100} (2006) 111.
%\bibitem{Paper2} B. Author1 et al., Phys. Rev. B {\bf 75} (2007) 222.


\bibitem {mott74}N.F. Mott, \textit{Metal-insulator Transitions} (Taylor and
Francis, London, 1974).

\bibitem {anderson58}P.W. Anderson, Phys. Rev. \textbf{109} (1958) 1498.

\bibitem {dmftrev}A. Georges \textit{et al.}, Rev. Mod. Phys. \textbf{68}
(1996) 13.

\bibitem {tmt}V. Dobrosavljevi\'{c} \textit{et al.}, Europhys. Lett.
\textbf{62} (2003) 76.

\bibitem {vollhardt}K. Byczuk, W. Hofstetter, and D. Vollhardt, Phys. Rev.
Lett. \textbf{94} (2005) 056404.

\bibitem {impurity}M.C.O. Aguiar \textit{et al.}, Phys. Rev. B \textbf{73}
(2006) 115117.

\bibitem {sb4}G. Kotliar and A.E. Ruckenstein, Phys. Rev. Lett. \textbf{57}
(1986) 1362.

\bibitem {screening}D. Tanaskovi{\' c} \textit{et al.}, Phys. Rev. Lett.
\textbf{91} (2003) 066603.


\end{thebibliography}
\end{document}